\newcommand{\solm}{M$_{\odot}$\ }

\newcommand{\rf}{\par\noindent\hangindent 15pt {}}

\documentclass[preprint2]{aastex}

\shorttitle{Dust Embedded GC Sources}
\shortauthors{Moultaka et al.}

\begin{document}


\title{M-Band Spectra of Dust Embedded Sources at the Galactic Center}


\author{J. Moultaka\altaffilmark{}}
\affil{Laboratoire d'Astrophysique de Toulouse-Tarbes, Universit\'e de Toulouse, CNRS, 14, Avenue Edouard Belin, 31400 Toulouse, France}
\email{jihane.moultaka@ast.obs-mip.fr}

\author{A. Eckart\altaffilmark{}}
\affil{I.Physikalisches Institut, Universit\"at zu K\"oln,
             Z\"ulpicher Str.77, 50937 K\"oln, Germany\\
       \& Max-Planck-Institut f\"ur Radioastronomie,
             Auf dem H\"ugel 69, 53121 Bonn, Germany}

\and

\author{R. Sch\"odel\altaffilmark{}}
\affil{Instituto de Astrof\'isica de Andaluc\'ia,
         Camino Bajo de Huétor 50, 18008 Granada, Spain\\
      \& I.Physikalisches Institut, Universit\"at zu K\"oln,
             Z\"ulpicher Str.77, 50937 K\"oln, Germany}




\begin{abstract}
The goal of the present paper is to investigate the circumstellar material around the brightest 
dust-enshrouded sources in the central stellar cluster of the Milky Way.
Observations have been carried out at the European Southern
Observatory's Very Large Telescope on Paranal, Chile. 
We have used the long wavelength (LWS3) low resolution (LR)
spectroscopic mode of the ISAAC camera at the VLT
in the spectral range of the M filter from 4.4$\mu$m to 5.1$\mu$m.
The use of a slit width of $0.6\arcsec$ implied a spectral resolution of $R= \lambda/\Delta \lambda$=800 ($\Delta v=375$~km/s).
These observations resulted in M-band spectra
of 15 bright sources in the central stellar cluster of the Milky Way.
In addition to gaseous $^{12}$CO (4.666$\mu$m) and $^{13}$CO (4.77$\mu$m) 
vibration-rotational absorptions,
we detect a strong absorption due to a mixture of
polar and apolar CO ice (centered at 4.675$\mu$m).
In the shorter wavelength absorption wing there is an absorption feature due to XCN at 4.62$\mu$m. 
The XCN absorption is strongest toward the M2 supergiant IRS7.
We find that the extinctions due to material traced by the CO ice and the CO gas absorptions may be of comparable importance.
Using the spectra of IRS~2L and IRS~16C we perform a first order correction of the
line of sight absorption due to CO-ice and $^{13}$CO gas. 
In combination with published hydrogen number density estimates
from sub-mm CO(7-6) and FIR [OI] line data we obtain gas masses of the circumstellar shells 
of the order of 10$^{-3}$ and 10$^{-2}$\solm.
This implies that in future spectra taken at high spectral and angular resolution 
the bright and dust embedded Galactic Center sources should show a substantial 
line absorption due to source intrinsic absorption. 
\end{abstract}





\section{Introduction}

At a distance of 8 kpc, the Galactic Center (GC) 
is the closest galactic central region. It can thus be studied in detail by 
the direct observation of individual stars, gas and dust. It represents an ideal
and unique case allowing one to analyze thoroughly the direct environment of a central 
super-massive black hole.\\
The Galactic Center is obscured mainly by extinction from the diffuse 
interstellar medium (ISM) present along the line of sight. About less than one third of this extinction arises from the dense foreground ISM (molecular clouds). The Galactic Center is surrounded by a 
clumpy, circum nuclear ring of dense gas and dust (G\"usten et al. 1987).
The exact magnitude and spatial distribution of this obscuration are still unclear 
but a mean value of the visual extinction towards prominent sources within the
central stellar cluster reaches $\sim 27$mag (e.g.  Sch\"odel  et al. 2007, Scoville et al. 2003, 
Lebofsky, Rieke, Tokunaga, 1982).
Moreover, the extinction across the central 10" to 20" is shown to be 
smoothly distributed at an angular resolution of 2'' 
(Sch\"odel  et al. 2007, see also Scoville et al. 2003).

In a previous work (Moultaka et al. 2004, 2005), we presented a spectroscopic study of the distribution of the 
water ice and hydrocarbon features at 3.0$\mu$m and 3.4-3.48$\mu$m
towards stars in the central $0.5$~pc of the GC stellar cluster that are
bright in the 2-4 $\mu$m wavelength domain.  
With the aid of additional
K-band spectroscopic data, we derived optical depth spectra of the
sources after fitting their continuum emission with a single reddened
blackbody continuum.  As a novel approach, in Moultaka et al. (2004), we 
also derived intrinsic source spectra by
correcting the line of sight extinction via the optical depth spectrum
of a GC late type star that is most likely not affected by local dust
emission or extinction at the Galactic Center. The good agreement
between the two approaches shows that the overall variation of the
line-of-sight extinction across the central $0.5$~pc is 
$\Delta A_{\mathrm{K}}\leq0.5$~mag. 
This is confirmed by a recent extinction map by Sch\"odel et al. (2007).
The extinction corrected spectra of the
hot He-stars are in good agreement with pure Rayleigh Jeans continuum
spectra.  The intrinsic spectra of all other sources show residual
dust continuum emission and/or grain mantel absorption features.
Since it is unlikely that foreground extinction patches are exclusively associated 
with the extended bright GC MIR sources we interpreted both facts as evidence that
a significant amount of the absorption takes place within the central
parsec of the Galactic Center and is most likely associated with  
the circum-stellar material in which the individual sources are embedded.

The goal of the present paper is to use the 4.55-4.80$\mu$m M-band spectra for several
bright sources in the central cluster \footnote{Resulting from ESO VLT observations of program ID number 75.C-0138A}
in order to 
distinguish between line 
of sight and possible source intrinsic absorption.
Prominent line features in the M-band spectrum are the Pf${\beta}$ hydrogen 
recombination line and absorption lines from gaseous and solid CO.

Cold gaseous CO produces the typical vibration-rotation 
band structure of a diatomic molecule, two broad envelopes of 
absorption separated by a small gap at 4.666$\mu$m (the $v=1-0$ band center; Allamandola 1984).
In addition to the rotational vibrational P and R branch absorption of 
gaseous $^{12}CO$ there is a $^{13}CO$ gas absorption band at 4.77$\mu$m 
and an absorption feature at 4.675$\mu$m attributed to solid phase CO.
Based on ISO SWS/LWS measurements 
Moneti, Cernicharo \& Pardo (2001) infer a strong gaseous CO absorption due to 
material with a bulk kinetic temperature of $\sim 10$K towards SgrA*.

Solid CO has a single narrow absorption line at 4.675$\mu$m ($\sim$0.01$\mu$m FWHM). 
In the presence of impurities the line width tends to increase 
and the central wavelength frequency tends to decrease
(see Hagen, Allamandola \& Greenberg 1979, 1980).
The width of the 4.675$\mu$m solid CO absorption is indicative of typical
concentrations and types of impurities present in the solid (Allamandola 1984).
The first detection of interstellar solid CO at 4.675$\mu$m was obtained in 
the deeply embedded young stellar object W33A (Lacy et al. 1984).
Studying the CO ice absorption towards proto-stellar objects 
Lacy et al. (1984) also found a broad absorption feature at 4.62$\mu$m.
The authors suggest that this XCN feature is due to UV-photolysis 
of cold grain mantles containing molecules with C-N bonds.
Laboratory spectra show that the changes in the band strengths are 
due to variations in the composition.

There are a number of detailed M-band spectroscopic studies on individual sources 
in the central stellar cluster.
Under the assumption that all of the solid-phase $4.675\mu$m CO absorption is due to 
foreground material, its detection towards IRS~12 and its absence 
towards IRS~3 and IRS~7, lead McFadzean et al. (1989) to the conclusion 
that the distribution of cold molecular clouds is not uniform. 
They also conclude that it is unlikely that the molecular clouds are 
close to the Galactic Center because the CO ice is unlikely to survive 
temperatures higher than 20~K unless it is located in an environment 
of water ice and other molecules.
CO gas-phase absorption has been observed in 
IRS~3, IRS~7 and IRS~12 by Geballe (1986) and McFadzean et al. (1989), 
but the P and R branches were not resolved
in these low resolution spectra.

Through Fabry-Perot spectroscopy at a spectral resolution of 20~km/s
and 40~km/s Geballe, Baas \& Wade (1989) have shown that a substantial fraction of the
M-band gaseous CO absorption is likely to be due to intervening spiral arms,
local molecular gas and clouds that 
are thought to be very close to the Galactic Center.
The two most prominent likely absorbers are the  20~km/s and 50~km/s clouds.
Measurements of the CO R(2) and R(5) lines towards IRS~1, 2, 3, 5, 6, 7 and 8
reveal that most of the absorption takes place in the velocity
interval between 0 and 75 km/s.

In this contribution, we present recent observations of the central parsec of
the Galaxy, performed in the M-band spectral domain. Our observations consist on a larger number of sources in the region (15 sources) than already done and they were performed at a higher spectral 
(except for the few sources studied with the Fabry-Perot
 spectrometer by Geballe, Baas \& Wade 1989)
and angular resolution. 
This allows us to analyze the spatial distribution
of the CO absorption in its gas- and solid-phase and a first order 
distinction between the amount of line of sight and 
possible source intrinsic absorption.

\section{Observations and data reduction}\label{sec:obs}

We observed the central parsec of the 
Galactic Center using the ISAAC spectrograph operating at the 
Antu UT1 unit of the ESO Very Large Telescope (VLT) located 
at the Paranal mountain in Chile. 
The observations were done in July 2005 during two allocated nights.

\begin{figure}
\includegraphics[width=9cm,angle=-00]{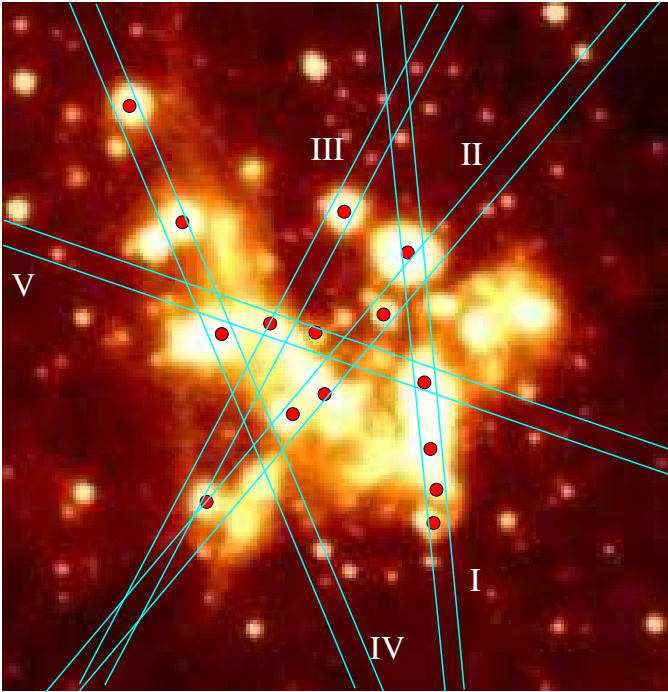}
\caption{\label{slit} 
Slit settings that we used to obtain M-band spectra of 
individual sources in the central stellar cluster.
The size of the ISAAC M-band image is 26''$\times$26''.
North is up and east is to the left.
The individual program sources are marked with a red dot.
From North to South the different slits contain the following sources:
slit {\bf I.}: IRS~3, IRS~13, IRS~2L, IRS~2S, IRS~12N;
slit {\bf II.}: IRS~3, IRS~29, IRS~16SW, IRS~21, IRS~9; 
slit {\bf III.}: IRS~7,  IRS~16NE, IRS~9; 
slit {\bf IV.}: IRS~5, IRS~10W, IRS~1W;  
slit {\bf V.} from East to West: IRS~16NE, IRS~16C, IRS~13.
}
\end{figure}

We have used the long wavelength (LWS3) low resolution (LR) spectroscopic mode of ISAAC 
in the M-band spectral filter range from 4.4$\mu$m to 5.1$\mu$m wavelength. 
With the combination of a $0.6\arcsec$ slit width, the measurements resulted in a spectral 
resolution ($R= \lambda/ \Delta \lambda$) of 800 ($\Delta v=375$~km/s). 
The slit height is of $120\arcsec$. 
The seeing varied from 0.7$\arcsec$ to 
2.7$\arcsec$ during the observing run.
The different slit settings that were used are shown in Fig.\ref{slit}.

In order to correct for the sky emission, the chopping technique was used with chopper 
throws of $\sim 20\arcsec$ along the slit. A chopped frame contains a positive trace 
image and a negative one. 
Two consecutive chopped frames present shifted image positions where the positive 
trace image of the first frame is at the same position as the negative one of the second frame. 
Such consecutive frames were then subtracted from each other to provide a single 
frame containing two negative trace images and a positive one with twice the 
intensity of the negative images.
\\
All array images were also divided by the flat fields and corrected for cosmic 
rays and for dispersion related distortions. 
The wavelength calibration was performed using the Xenon-Argon lamps 
and the grating in 3rd order. 
Finally, all extracted spectra were corrected for telluric lines using 
different standards: HR6070 (A0V), HR6537 (A0V), HD169588 (A0V).
The standard star spectra show a weak Pf$\beta$ line that was not corrected for.
However, since we restricted  ourselves to standard stars
of the same spectral type and luminosity (i.e. AOV), a {\it relative} comparison
of the Pf$\beta$ line strengths between the resulting GC source spectra may be
done safely.

\section{Results}\label{Results}

We obtained high quPality M-band spectra of 15 bright sources of 
the central parsec including the hot (and intrinsically dust free) He emission-line stars 
of the IRS~16 complex (IRS16SW, 16NE, and 16C), 
cool stars and the dust-embedded sources of the northern arm of the mini-spiral. 
Four of these objects are bow shock sources that interact with the GC interstellar medium in the
northern arm (IRS~5, 10W, 1W, 21; Tanner et al. 2002, 2005, Ott, Eckart, Genzel, 1999).
IRS~3 and IRS~7 are known to interact
with a combination  of the wind from the hot He-stars and  a possible wind from SgrA*
(Viehmann et al. 2005, 2006, Yusef-Zadeh \& Morris 1991, Muzic et al. 2007).
IRS13 is a cluster of luminous young and hot stars (e.g. Paumard et al. 2006).
The sources IRS~12N, 9, 2S are late type stars.
IRS~2L is likely a B-type star (Cl\'enet et al. 2001). 
IRS~29 may have contributions from both IRS~29N (early type WC9) and IRS~29S (late type).
For references concerning the GC sources
see e.g. Genzel et al. (2000), Paumard et al. (2006) and
Viehmann et al. (2005; Fig.2 therein).

As it can be seen in figures \ref{IRS16}, \ref{coolstars}, \ref{dustembedded} and \ref{peculiar}, 
in addition to the Pf${\beta}$ hydrogen line at 4.65$\mu$m, 
four spectral features 
are detected in most of the sources in this spectral region:

\begin{itemize}
\item The absorption bands of the $^{12}$CO and $^{13}$CO gas-phase with their fundamental 
vibration-rotation P and R branch lines around 4.666$\mu$m and 4.77$\mu$m, respectively
(see Fig.~\ref{icegas}b). 
\item The absorption feature at 4.675$\mu$m of the solid-phase CO.
\item A broad feature at 4.62$\mu$m wavelength that can possibly be identified with an XCN absorption.
\end{itemize}

Unlike in the lower resolution data presented by McFadzean et al. (1989),
in our spectra, the P and R branches of the gas-phase 
absorption feature are resolved. 
We also observe a prominent
CO solid-phase absorption feature in IRS~7 
that was not detected previously by the authors.
In our case, prominent contributions of solid CO ice to the 
absorption features are
shown in IRS~2S, 2L, 12N, 7, 9, 21 and possibly in IRS~13 and 16SW.
Contributions of $^{12}$CO gaseous absorption is seen at low levels in
the spectra of probably all sources. It is strongest present
towards dust embedded sources like IRS~1W, 3, 5, 10W, and also towards
IRS~12N, 9, 13, 29.
Contributions from $^{13}$CO are weaker but can be seen well in IRS~3
(Fig.~\ref{icegas}).
We see indications for strong residual CO gaseous absorptions
towards extended infrared excess sources and weak residual absorptions
toward more compact and hot stellar sources in the Galactic Center.
These two facts support the presence of source intrinsic CO absorption
associated with individual sources at the Galactic Center.

\begin{figure}
\includegraphics[width=9cm,angle=0]{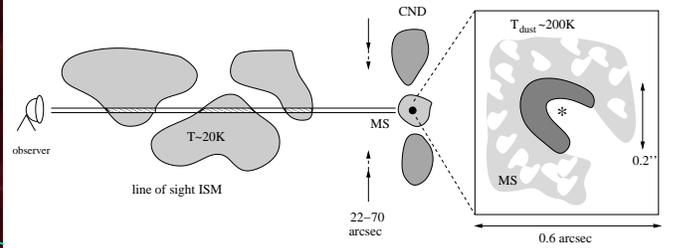}
\caption{\label{bild} Absorption along the line of sight towards the compact dust embedded 
sources in the central stellar cluster. In addition to the $\sim$20''-70''  ISO SWS/LWS apertures used by 
Moneti, Chernicharo \& Pardo (2001) to investigate the ISM via MIR line features in this direction,
we show the 0.6'' ISAAC M-band slit-width and a zoom towards a typical 0.2'' diameter bow shock like object in the
central few arcseconds - embedded in the mini-spiral material with a given area filling factor $\eta$. 
The central star is indicated by an asterix. The abbreviations MS and CND stand 
for mini-spiral and circum nuclear disk.
}
\end{figure}

With large apertures in the range of 20'' to 70'' as used by 
Moneti, Cernicharo \& Pardo (2001) the ISO SWS/LWS measurements are dominated by the absorption in 
the cold ISM along the line of sight against the strong thermal emission of the 
extended SgrA~West source, like the mini-spiral and the inner extensions
of the circum nuclear ring (CND). \\
While the contribution of compact absorption sources at the
Galactic Center is included in apertures of these sizes, their beam filling 
is small with respect to the almost totally beam filling  extended source components.
This  situation changes, however, if we go to apertures as small as the ISAAC slit-width of
only 0.6''.
In Fig.\ref{bild} we show the geometry that is relevant to understand the MIR absorption
line measurements towards sources in the Galactic Center. Here the flux density contribution of the compact dust emitting sources located at the Galactic Center
becomes increasingly dominant and the contribution of the much more extended flux density contributors
is strongly diminished.
In addition to the absorption due to the cold material along the line of sight
towards the Galactic Center, the source intrinsic absorptions now become increasingly 
important.
In the case of the present sources the absorption occurs within the circumstellar gas and 
dust shell against a number of background sources: 
1) the stellar emission, 
2) the thermal emission of the mini-spiral background,
3) and against a combination of the strong contribution of scattered light and thermal emission 
of the circumstellar shell itself.

\subsection{The Foreground absorption}
In the present section we estimate the line of sight foreground 
extinction due to CO ice and gas absorption (the latter, based on $^{13}$CO gas line measurements). After comparison with the absorption towards an
intrinsically dust and molecular gas free GC source the residual CO absorption toward other 
GC sources in the field can then be attributed to the source intrinsic 
extinction 
under the assumption that the foreground extinction does not 
vary on scales of about 20". 
In the following, the optical depth is calculated via
$\tau = - ln(I_a/I_c)$, where $I_a$ and $I_c$ are the line and continuum intensity at a given wavelength. The continuum level is determined 
for all sources as the best fitting straight line through the
intensities at wavelengths less than 4.56$\mu$m (except for IRS16C where we took wavelengths between 4.56 and 4.58 micron because the spectrum is not reliable at shorter wavelengths), larger than 4.80$\mu$m, 
and through a 0.01$\mu$m wide interval centered at 4.73$\mu$m, 
i.e. between the $^{12}$CO and $^{13}$CO gas absorption.

\subsubsection{Calibrating the ice absorption}
\label{Ice}
In order to calibrate the CO ice 
we selected IRS2L as a calibration source. Indeed, IRS2L has been identified as a B-type star (Cl\'enet et al. 2001), it is thus free of CO bandhead absorption and has a relatively flat spectrum in the M-band. Its spectrum shows a high signal to noise and well pronounced ice absorption with only weak traces of CO-gas absorption. 
Moreover, we have assumed that the overall absorption towards IRS~2L is dominated by line of sight ice-absorption since this source is located close but not inside the mini-spiral area and is clearly not behind the CND material. For all these reasons, the line shape of the IRS2L spectrum can be used as a template for the foreground ice absorption towards all other sources in the field and to correct for the contribution of this ice-feature (see Figs.~\ref{coolstars} and \ref{icegas}a). In the following, we assume that the foreground CO ice absorption has the shape of the template spectrum but we allow for a small variation in the amount of this absorption from source to source.\\
On the other hand, the absorption trough towards short wavelengths in the IRS2L spectrum is indicative of a possible contribution of XCN absorption (see section \ref{xcn} and references therein). Therefore, this spectrum is also well suited to calibrate a possible XCN line of sight absorption.
\\

\begin{figure}
\includegraphics[width=9cm,angle=0]{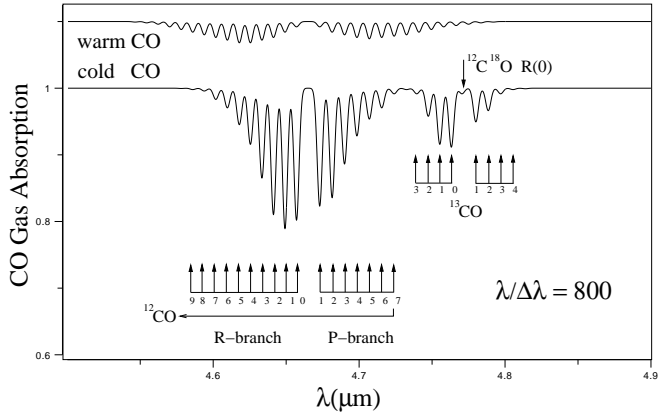}
\caption{\label{COtheofig} M-band 4.666$\mu$m spectra of the rotation-vibration modes of CO for the case of
warm (T$\sim$150~K) and cold (T$\sim$10~K) 
gas as published by Moneti, Cernicharo \& Pardo (2001) but here shown at a
spectral resolution of $\lambda/\Delta\lambda$=800 ($\Delta v=375$~km/s) as achieved with the ISAAC spectrometer.
At this spectral resolution the individual vibration-rotational lines cannot be fully separated from each other
and therefore form a through like overall absorption over the R- and P-branch wavelength range.
The absorption free section between the two bands, however, remains at the continuum level 
for this spectral resolution - a feature that is used to perform  the CO ice calibration 
described in section~\ref{Ice}.
The warm gas spectrum has been shifted up by 0.1 for display purposes.
Labeled are the ground level J values for the R- and P-branches with
$\Delta$J=+1 and $\Delta$J=-1, respectively.
Column densities are of the order of N($^{12}$CO)=10$^{17}$cm$^{-2}$ for the warm and 
a few times 10$^{18}$cm$^{-2}$ for the cold CO (Moneti, Cernicharo \& Pardo 2001).
}
\end{figure}

\begin{figure}
\includegraphics[width=8cm,angle=-00]{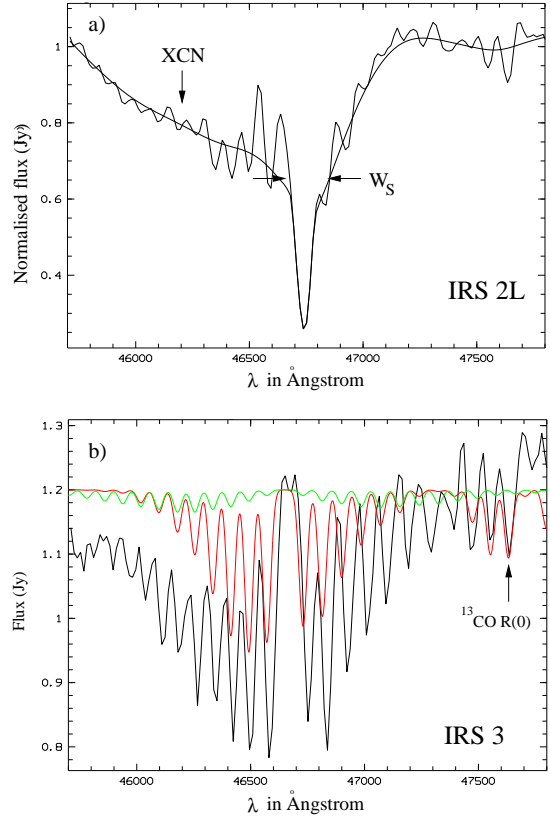}
\caption{ 
{\bf a)}
The CO ice absorption derived after smoothing the spectrum of 
IRS~2L (superimposed) and continuum normalised to one. The ice absorption has a FWHM of $W_S$.
This spectrum was used to correct the spectra of all sources.
For comparison with the CO ice corrected spectra see 
Fig. \ref{extcorr}.
{\bf b)} 
Comparison of the CO ice corrected M-band spectrum of 
IRS~3 to the theoretical spectra of the stretching modes of gaseous CO 
shown in Fig.\ref{COtheofig} (warm CO in green, cold CO in red, IRS~3 data in black). 
The theoretical spectra have been scaled to the
continuum level of the IRS~3 spectrum at 4.730 $\mu$m between the 
$^{12}$CO and the weak $^{13}$CO absorptions.
The comparison indicates the presence of the $^{13}$CO R(0), R(1) and R(2) lines
with a signal to noise of 3-5, given possible baseline uncertainties.
The fact that the $^{13}$CO lines appear to be a bit stronger compared to the 
model we attribute to noise and mostly to the baseline uncertainties.
}
\label{icegas} 
\end{figure}

\begin{figure*}
\begin{minipage}{38pc}
\includegraphics[width=27pc,angle=90]{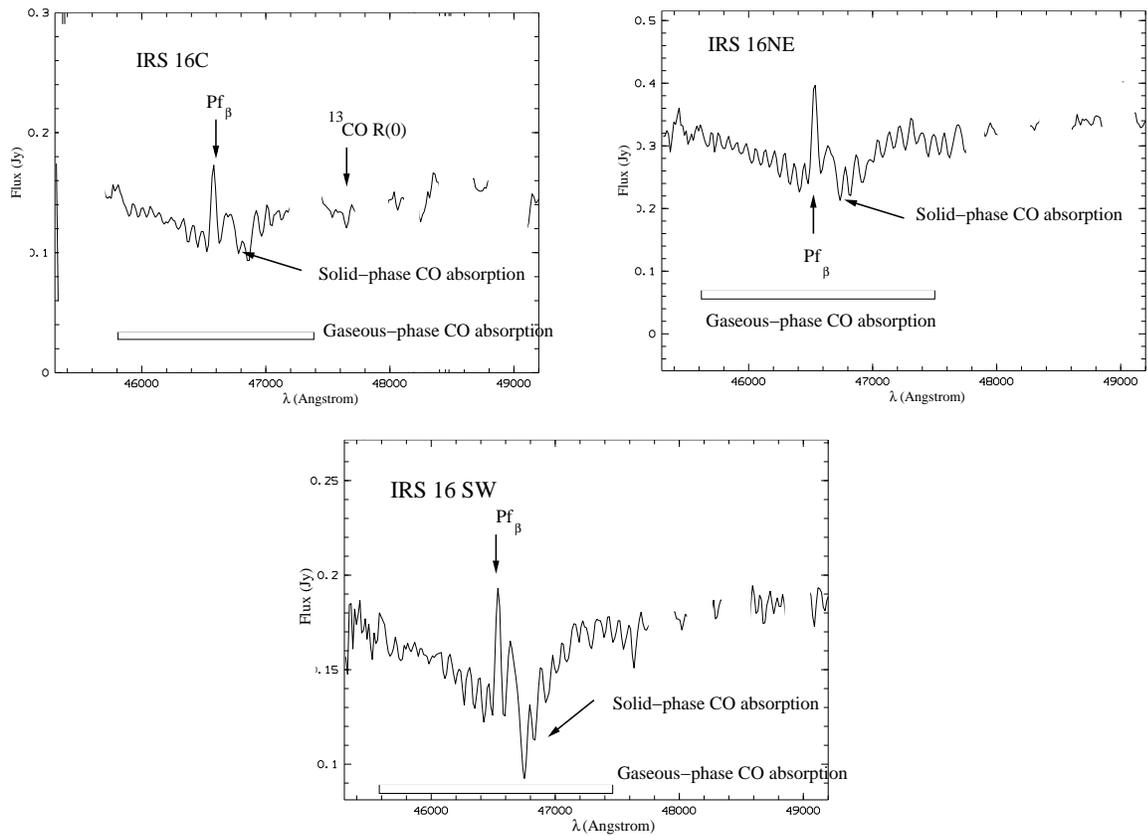}
\caption{\label{IRS16} M-band spectra of the He emission line stars of the IRS~16 
complex. The spectra show CO gas- and solid-phase absorptions. All blanked 
spectral regions correspond to residuals from the telluric lines.}
\end{minipage}
\end{figure*}

\begin{figure*}
\begin{minipage}{38pc}
\includegraphics[width=27pc,angle=90]{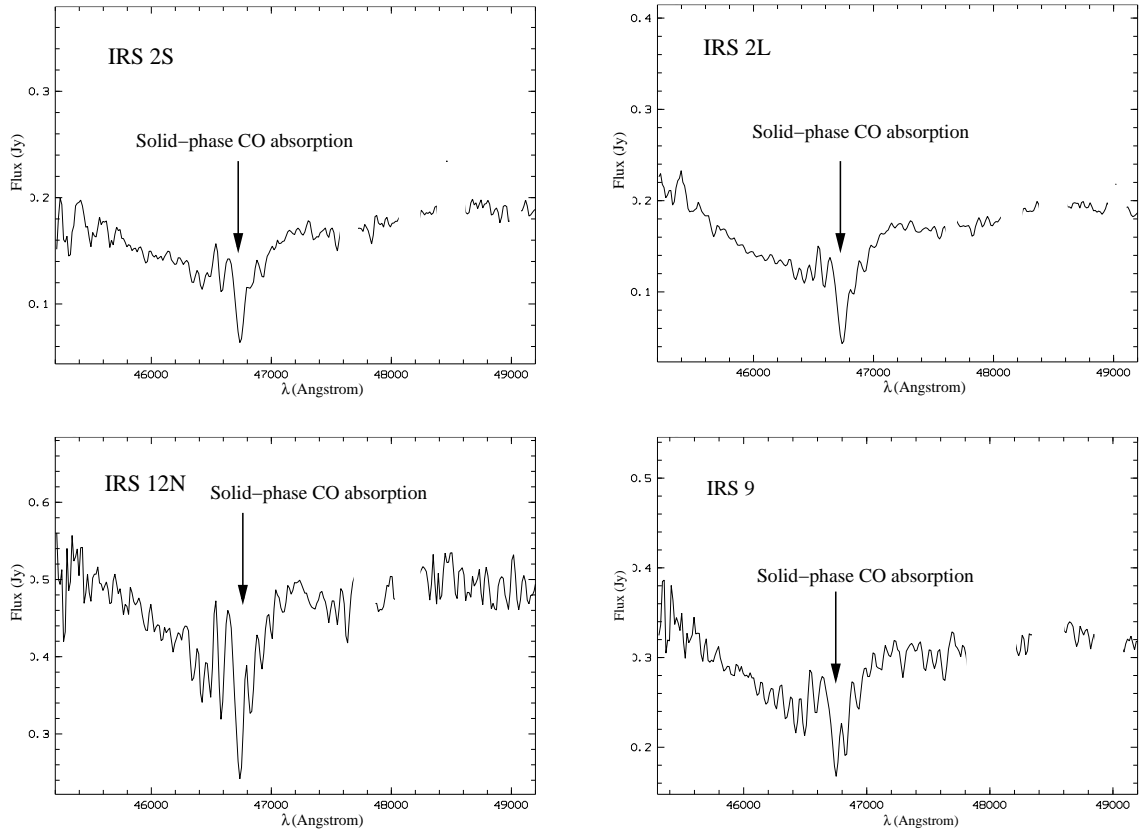}
\caption{\label{coolstars} 
M-band spectra of cool stars at the Galactic Center.
All of the sources shown in this figure are identified as late-type,
with the exception of IRS2L, that shows a featureless NIR spectrum. 
All blanked spectral regions correspond to residuals from the telluric lines.
}
\end{minipage}
\end{figure*}

\begin{figure*}
\begin{minipage}{38pc}
\includegraphics[width=27pc,angle=90]{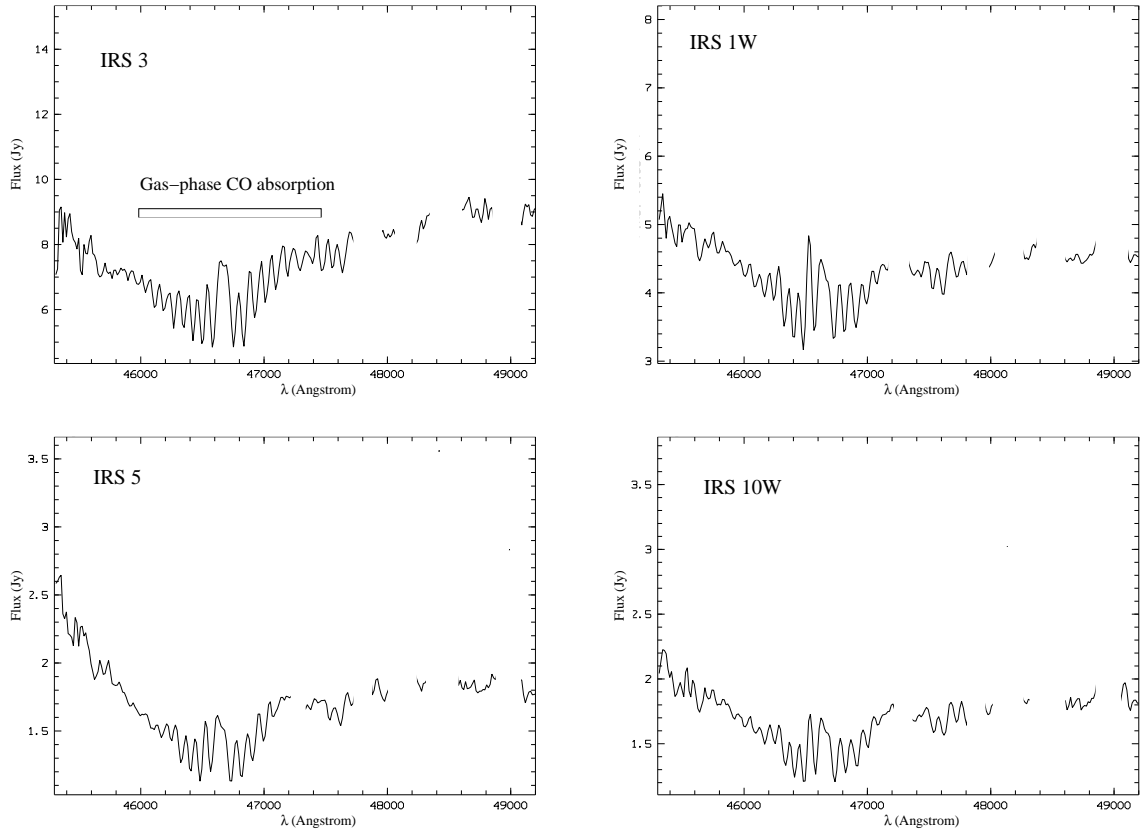}
\caption{\label{dustembedded} M-band spectra of the dust embedded sources IRS~3 and 
bright sources IRS~1W, IRS~5, and IRS~10W located in the norther arm of the mini-spiral. 
All blanked spectral regions correspond to 
residuals from the telluric lines.}
\end{minipage}
\end{figure*}

\begin{figure*}
\begin{minipage}{38pc}
\includegraphics[width=27pc,angle=90]{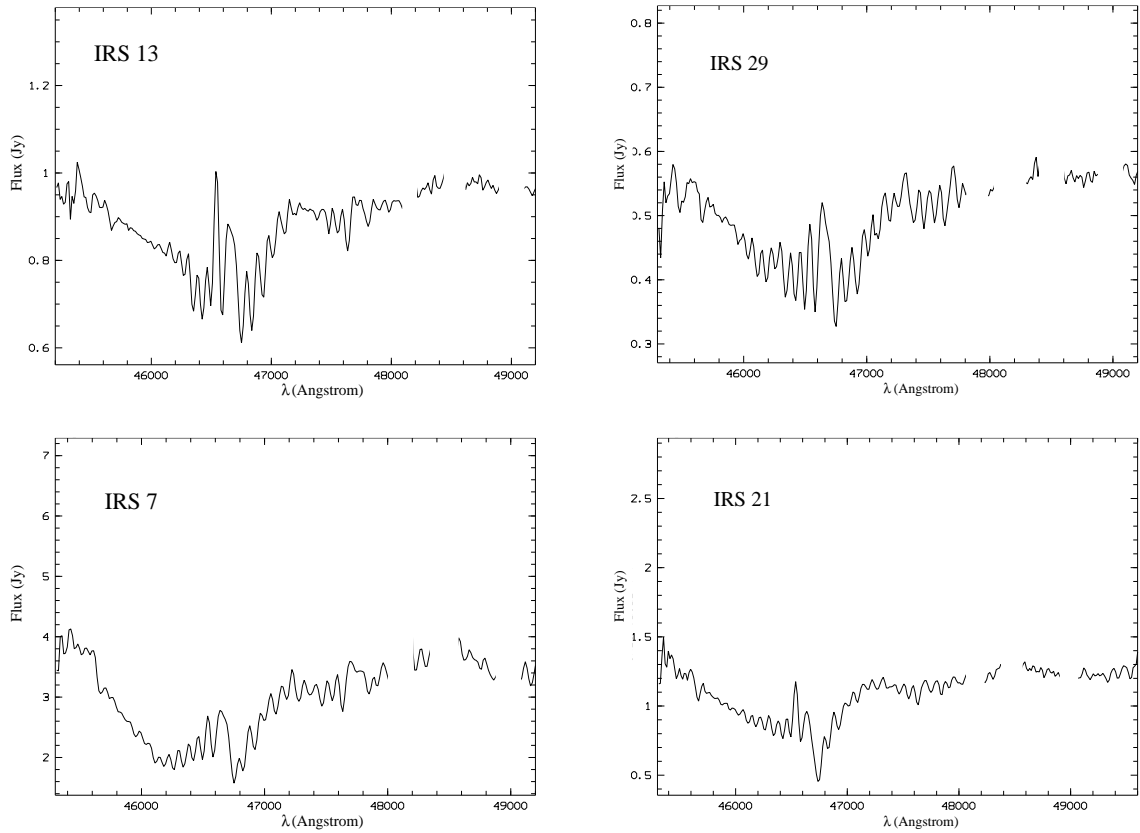}
\caption{\label{peculiar} M-band spectra of the IRS~13 complex, IRS~29, the dust
 embedded source IRS~21 and the cool supergiant IRS~7. All blanked spectral 
regions correspond to residuals from the telluric lines.}
\end{minipage}
\end{figure*}

To derive the CO ice absorption spectrum shown in Fig. \ref{icegas}a, we smoothed the IRS~2L spectrum and got rid, in this way, of the features not belonging to the CO ice feature, which are the Pf$_\beta$ and the prominent feature between the P- and R- branches (at about 4.67 $\mu$) of the CO gas feature; we then normalised the continuum to one. \\
The individual source spectra were divided by the resulting ice foreground absorption spectrum diluted by an additional adjustable continuum contribution (which accounts for a possible variation in the amount of absorption), with the condition that the spectral region between the 
P- and R-branches of the remaining gaseous absorption should be at the same flux density level as the
continuum around 4.73$\mu$m and 4.80$\mu$m wavelength.\\  
In this process, we assumed that the spectral region 
centered at a wavelength of 4.73$\mu$m and 4.80$\mu$m (i.e. between the $^{12}$CO and $^{13}$CO
and longward of the $^{13}$CO absorption features) is a continuum and therefore, must be at the 
same level as the continuum region longward of the $^{13}$CO absorption feature in the final 
corrected spectra as it is the case in the theoretical CO spectrum (see Figs. \ref{COtheofig} 
and \ref{icegas}b). 
This procedure is feasible since both the CO ice absorption and the center of the 
CO gas absorption are well separated from each other (by about a FWHM of the CO ice absorption feature).\\
The resulting spectra are shown in Fig. \ref{extcorr}.

Under these conditions we find a mean value of the diluting continuum of about 4~Jy for all the sources in the field, including the calibrator source IRS~2L (the values of the diluting continuum toward individual sources are shown in Tab.\ref{dilution}). This means that the mean foreground solid CO extinction spectrum is obtained by shifting the template spectrum shown in Fig.4a by 4~Jy. Then, the continuum is at 5~Jy (4+1~Jy) and the minimum of the absorption is at about 4.25~Jy (4+0.25~Jy). Defining the ``absorption strength'' as being the ratio between the continuum intensity and the minimum absorption intensity (i.e; I$_c$/I$_a$),  the solid CO absorption strength is about 1.2 (i.e. 5/4.26$\sim$1.2), while the one of the template is of 3.8 (1/0.26$\sim$3.8). This means that about 1/3 of the ice template absorption strength is due to foreground material along the line of sight which corresponds to an optical depth of $\tau$$\sim$0.1-0.2 ($\tau$=-ln(I$_a$/I$_c$)). For residual ice-absorption see comments in section \ref{section:Notes}. Allowing for larger amounts of foreground absorption leads to an over-correction of the spectral region between the CO P- and R-branches.\\
Please notice that a dilution was also needed in the case of the IRS~2L spectrum (which was used to derive the template spectrum) because gaseous CO and Pf$_\beta$ features are also present in this spectrum even though they are not very prominent. Thus, correcting this spectrum for the foreground solid CO extinction should also take into account that the region between the P- and R- branches of the gaseous CO feature and the regions between the $^{12}$CO and $^{13}$CO and longward of the $^{13}$CO absorption features are part of the continuum.\\

\subsubsection{The extinction associated with the CO ice absorption}\label{Gas}
The column density of solid-state CO can be derived via
N(CO)$_s$$\sim$$\tau$$W_S$/A
with the FWHM of the line $W_S$ in cm$^{-1}$ and with the absorption strength
A=1.7$\times$10$^{-17}$cm molecule$^{-1}$
(Sandford et al. 1988). 
From the fact that on average about one third of the ice-template depth (Fig.\ref{icegas}a)
absorption is due to foreground material 
in order to reproduce the line shape of the CO gaseous absorption shown in
Fig.\ref{COtheofig},
we derive an optical depth of the foreground solid CO absorption of
$\tau$$\sim$0.1-0.2. 
With $W_S$$\sim$6-7~cm$^{-1}$$\sim$0.016$\mu$m (see Fig.\ref{icegas}) we obtain
N(CO)$\sim$7$\times$10$^{16}$cm$^{-2}$
(see Tab.~\ref{abundance}).

Using a CO versus H$_2$ abundance ratio for typical diffuse Galactic molecular clouds
([$^{12}$CO]/[H$_2$]=3$\times$10$^{-6}$), we then estimate the 
molecular gas column density associated to the foreground solid CO as N(H$_2$)$\sim$2.3$\times$10$^{22}$cm$^{-2}$ (see Tab.~\ref{abundance} for more details).
If the absorption takes place in the diffuse interstellar medium we find 
values for the visible extinction of A$_V$=14-28 (for $\tau$=0.1) or 27-53 (for $\tau$=0.2) using the value of A$_V$/N(H$_2$) of about 1-2$\times$10$^{-21}$mag cm$^{-2}$ 
(Bohlin et al. 1978, Moneti, Cernicharo \& Pardo 2001)
(see Tab.~\ref{abundance}). An A$_V$/N(H$_2$) of 1$\times$10$^{-21}$mag cm$^{-2}$ would be consistent with the case where a dominant fraction of neutral hydrogen is molecular (Bohlin et al. 1978).

In several nearby galaxies (e.g. M33, NGC4565, M101) far-infrared continuum measurements
have revealed the presence of an extended T=16-18~K cold dust component that is smoothly
distributed over the disk and is likely heated by the diffuse interstellar radiation
(Neininger et al. 1996, Hippelein, et al. 2003, Suzuki et al. 2009).
Warmer dust components are more concentrated towards the spiral arms of these systems
and associated with star forming regions. 
The temperature range is right at or just below the temperatures at which
CO desorbs from ice mantels, such that one can expect an appreciable amount of CO-ice
being associated with this component
(Collings et al. 2003, Bisschop et al. 2006, Acharyya et al. 2007).
This ISM phase can obviously not exclusively be associated with the dense 
star forming molecular cloud complexes in the spiral arms but covers an extended cold 
dust distribution also strongly associated with the inter-arm region.
It is quite likely that a similarly extended component is existent within the Milky Way,
such that the line of sight towards the Galactic Center samples this cold ISM phase 
extensively.

Low temperatures (Burgh, France \& McCandliss, 2007) 
and the relevance of hydrogenation processes on grain surfaces 
(O'Neill, Viti \& Williams, 2002, Viti, Williams \& O'Neill, 2000)
indicate that in fact a very large portion of CO may be frozen out
in diffuse and translucent clouds.
This indicates that a substantial fraction of
the extinction is caused by the dust associated with the corresponding 
more volatile H$_2$ gas that remains in the gas phase 
(see e.g.  Duley 1974, Lee 1972, 1975).
Alternatively, mixed models in which a significant portion of the
extinction takes place in a diffuse ISM, with little
CO and no CO ice, and in dense clouds 
with higher abundances of CO and CO ice are plausible as well.

This indicates that the hydrogen column (along with the dust) associated with the
frozen out CO can fully 
account for the extinction of 25 - 30 mag towards the Galactic Center
which is consistent with the findings by Sch\"odel  et al. (2007), Scoville et al. (2003).
In case that the absorption takes place predominantly in dense molecular 
clouds, we find values of A$_V$ =1-2 which do not account for the observed values of 25-30 mag. (see Tab.~\ref{abundance}). \\
Since only less than one third of the interstellar extinction 
can be attributed to the dense interstellar medium (Whittet et al. 1997), more than 2 thirds have to be attributed to the diffuse Galactic interstellar medium. If this latter is in a cold neutral phase (CNM) (Jones 2002, Wolfire et al. 1995, Field, Goldsmith, Habing 1969,
Ferriere, Zweibel, Shull 1988), it would support our finding and our use of the molecular abundance value of the diffuse ISM.

\subsubsection{The gaseous CO absorption} 
\label{The-gaseous-CO-absorption}
 
\begin{figure*}
\includegraphics[width=18cm,angle=0]{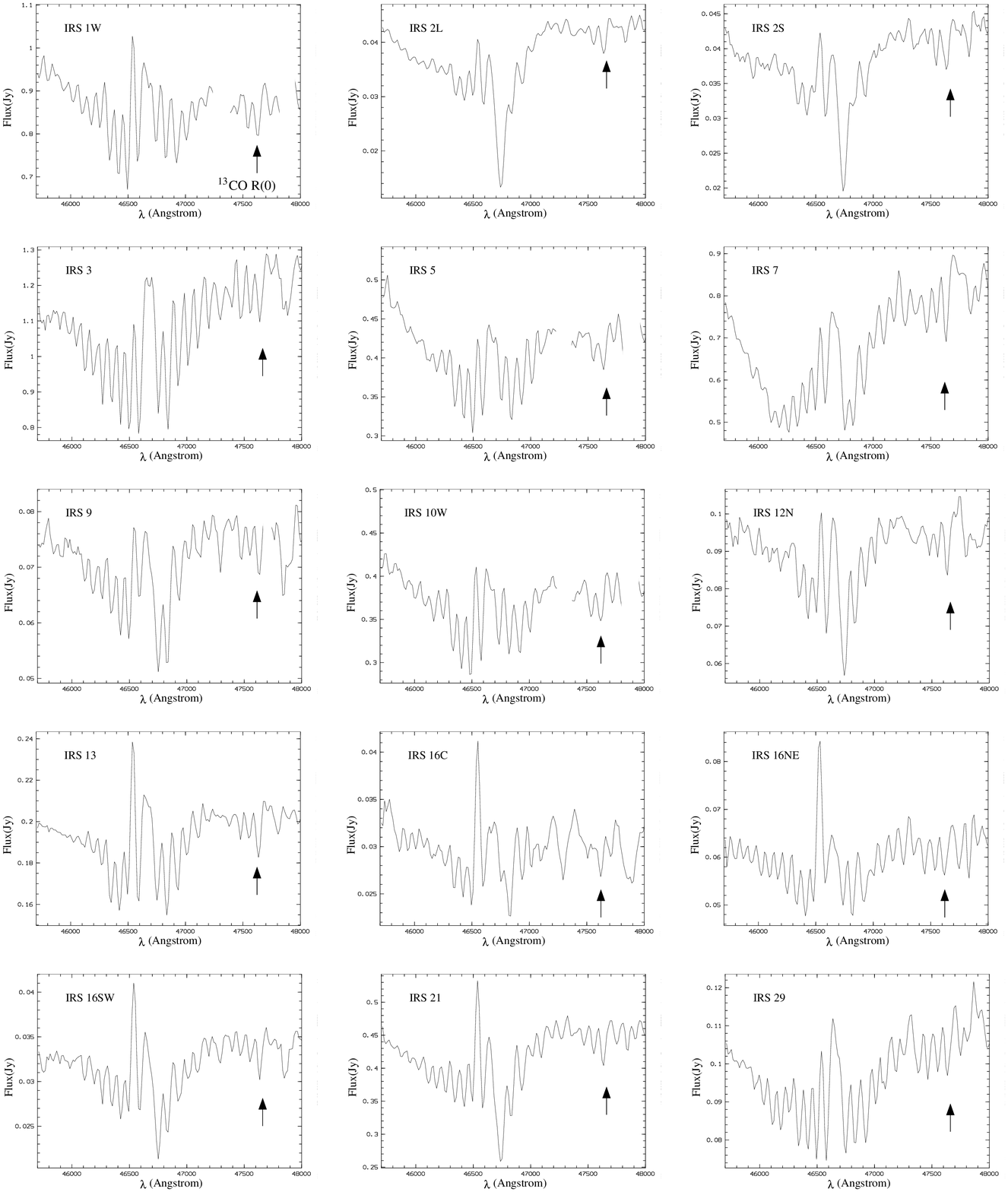}
\caption{Spectra of 15 Galactic Center sources corrected for foreground 
CO ice absorption. The position of the $^{13}$CO R(0) line is marked by arrows.
}
\label{extcorr} 
\end{figure*}

The R=800 ($\Delta v=375$~km/s) spectra corrected for CO ice absorption show a gaseous $^{12}$CO rotation-vibration absorption around $4.666\mu$m and an indication for a gaseous
$^{13}$CO rotation-vibration absorption around $4.77\mu$m. These absorption features are weak in the IRS2L spectrum, therefore the latter was not used to calibrate the foreground gaseous $^{13}$CO absorption. 
We show in 
Fig.\ref{COtheofig} the theoretical spectrum of these absorptions as calculated by Moneti, Cernicharo \& Pardo (2001) smoothed to our spectral resolution. 
Compared to this model, our observed relative strengths of the two isotopic line 
absorptions (see Fig.~\ref{extcorr})
indicate temperatures higher than T=10~K
or the presence of a mixture of warm (T$>>$10~K) and 
cold (T$\le$10~K) gas absorption.\\
As stated by Moneti, Cernicharo \& Pardo (2001), the $^{12}$CO lines are highly saturated (also, Geballe, Baas \& Wade (1989) quote optical depths of $\tau$$\sim$2 in the J=1-0 R(2) line of IRS~1 and IRS~3  with overall lower absorption depths in the J=1-0 R(5) line). On the other hand, due to the low abundance of the $^{13}$CO, its lines are not expected to be saturated. Therefore, in what follows, we will use the prominent $^{13}$CO R(0) line to estimate the foreground CO gaseous absorption and its corresponding optical extinction.\\
For the hot and non-embedded IRS16 sources the estimate of the $^{13}$CO
absorption strength corresponds to that expected for cold (T$\le$10~K) 
molecular gas (see Fig.~\ref{extcorr}).
This is also true for IRS16C that is located off the mini-spiral and
therefore also
most likely free of local GC absorption. Since the gas and dust temperatures at the Galactic Center are much higher than
that, we identify this $^{13}$CO line absorption as being due to foreground extinction due to
gaseous
and determine an optical depth at our spectral resolution of $\tau_{^{13} CO}$$\sim$0.11$\pm 0.02$.

Assuming a fractional abundance in dense molecular clouds of
[$^{13}$CO]/[H$_2$]$\sim$5~10$^{-6}$ (Rodriguez-Fernandez et al. 2000) and a temperature in the foreground material of the order of $\sim$10K, we get a $^{12}$CO column density based on the $^{13}$CO absorption of about N($^{12}$CO)$\sim$3.2~10$^{18}$, using the following equation (Geballe, Wollman \& Rank 1972,
Tielens et al. 1991,
Moneti, Cernicharo \& Pardo 2001,
Eyres et al. 2004): 
\begin{equation}
N(^{13}CO)_g \sim 3\times10^{14} \tau T \Delta v/(J+1) \times exp(E/kT)
\label{equ1}
\end{equation}
where J and E are the rotational quantum number and the energy 
above the ground of the absorbing level, respectively. Our result is consistent with 
the $^{12}$CO column density found by Moneti of N($^{12}$CO)$_g$$\sim$6.6$\times$10$^{18}$.

Given the uncertainties in the temperature, the isotopic CO abundance ratio, the $A_V/N(H_2)$ ratio, optical depth measurement and the spectral resolution, the corresponding visual extinction A$_V$ covers a large range of values with a minimum of the order of 25 mag which is consistent with previously published results. This indicates that the gaseous CO absorption in dense molecular clouds can also account for the foreground extinction to the Galactic Center and that it can be comparable to the amount of extinction traced by the deep CO ice absorption features in diffuse clouds. 
In chapter \ref{massestim}
we investigate how much of the extinction may be associated with the 
dust and gas envelopes around the embedded stars at the Galactic Center.

\subsection{Notes on individual spectra}
\label{section:Notes}

An inspection of the resulting  source spectra corrected for CO-ice absorption 
(shown in Fig.\ref{extcorr}) reveals residual contributions of CO-ice in IRS~2L, 2S, 7, 9, 12N, 21 and towards IRS16~SW.

This residual absorption may be due to variations in the foreground line of sight
extinction since the extinction in the central 10$\arcsec$ to 20$\arcsec$ is shown to be smoothly distributed only at a resolution of 2$\arcsec$ (see Introduction). In this case the absorbing material would not be located close to the Galactic Center.
Water ice survives at temperatures of about 100K, whereas CO will not stick on
water ice mantles at temperatures of $>60$K. Therefore the possible presence of H$_2$O ice 
(e.g. Moultaka et al. 2004, 2005) does not necessarily imply the presence of CO ice on the same grains.
Also the mean dust temperature within the mini-spiral is about 200~K (Cotera et al. 1999)
and there is no direct evidence for temperatures as low as 60~K in the vicinity of the central parsec.

It can, however, not be fully excluded that part of the residual solid-phase CO and possible XCN absorption 
in the corrected spectra shown in Fig.~\ref{extcorr} may in fact be due to 
the local Galactic Center material (i.e. mini-spiral) or to material within the circumstellar shells 
or bow shocks of the corresponding sources given above. 
This is consistent with the fact that the residual absorption appears 
to be narrower in the cases of IRS~2S and 12N.
Narrower CO-ice absorptions would be expected if the ice contains contributions of molecules with
low dipole moment (CH$_4$ of pure CO-ice) or high (T$\ge$150~K) deposition or desorption temperatures
(Sandford et al. 1988).

Bow-shock sources (e.g. Tanner et al. 2002, 2005), narrow dust
filaments (Muzic et al. 2007, Paumard et al. 2001, Morris 2000), and even the indication
for dust embedded young stellar objects (IRS13N -
Muzic et al. 2008) suggest the presence of high density pockets and high optical depths.
At about 300km/s the travel time through the central light year
is only of the order of 10$^3$ years matching the photo-evaporation timescales
of molecular clumps and disks in similarly harsh environments. These 
photo-evaporation timescales range
from 10$^3$ to 10$^5$ years (e.g. Mellema et al. 1998 van Loon \& Oliveira 2003).
This indicates that these compact dusty structures can persist while 
traveling through the central parsec.
At large optical depths (Av$>$4-6) the gas
temperature usually falls below the dust temperature
(e.g. Fong et al. 2001).
An average dust temperature of 200~K therefore does not exclude the presence of lower
temperature material.
This is consistent with the fact that Moultaka et al. (2004, 2005) 
find intrinsic water ice absorption towards all of these sources. Therefore the solid-phase 
CO can probably not be formed but may very well survive in the Galactic Center environment.

Towards all sources which are MIR bright and embedded (e.g. Viehmann et al. 2005, 2006),
the CO-gas absorptions are stronger - especially towards IRS~3, 29, 5, 1W and 10W than towards the non-embedded sources.
This suggests that part of this absorption is local and associated with these objects. The CO-gas absorption is 
lowest towards the late type sources IRS2L, IRS2S 
and the hot stars IRS16~NE, IRS16~SW and in particular IRS16C. 
In sections~\ref{Gas} and~\ref{massestim} 
we attribute the absorption towards IRS16C as line of sight absorption 
due to cold CO-gas.

The Pf$\beta$ line emission is stronger towards 
IRS~1W,
IRS~10W,
IRS~13,
IRS~21, and towards the 
IRS~16 sources.
These sources are located along the mini-spiral or are just off-set from it by an arcsecond or less.
Towards all other sources the Pf$\beta$ line emission is much weaker.
We conclude that the Pf$\beta$ line emission traces the ionized gas within the extended 
mini-spiral.

\subsection{The 4.62 $\mu$m XCN feature}\label{xcn}

A broad absorption feature centered at 4.62 $\mu$m is seen in the spectra
 of several YSOs and is attributed to triple CN bonds in 
nitrile or iso-nitrile ('XCN'; 
Lacy et al. 1984, Tegler et al. 
1993, 1995; 
Pendleton 1999, 
Chiar 1998,
van Broekhuizen et al. 2005).
 Laboratory studies have produced a similar feature by ultraviolet 
photolysis of CO and NH$_3$ 
(Lacy et al. 1984; 
Bernstein 1995; 
Schutte \& Greenberg 1997
) and assigned its carrier to OCN$^{-}$. 
Pendleton et al. (1999) explore several possible carriers of the 
XCN band and review possible production pathways through far-ultraviolet photolysis (FUV), 
ion bombardment of interstellar ice analog mixtures, and acid-base reactions.
W33A has the strongest XCN feature observed to date.
As XCN is thought to be the product of energetic processing 
of simple N-containing ices, the strong XCN feature 
is suggestive of a dust population that has been processed 
by irradiation, such as UV irradiation or ion bombardment (Lacy et al. 1984; Tegler 1993).

For IRS~7, the residual absorption at $\sim$4.62$\mu$m that may be associated with an
XCN feature (see Fig. \ref{extcorr}) is very pronounced while for all other 
sources it is only tentative like in IRS~2L, IRS~2S and IRS~5 or even absent. 
In our case, it is most probably associated to OCN${^-}$ (see van Broekhuizen et al. 2005).\\
Such a deep absorption band has already been reported for the spectra of IRS~19
(Chiar et al. 2002) and Sgr A* (Moneti, Cernicharo \& Pardo 2001). For Sgr A* the evidence for 
an XCN feature in the spectrum taken with ISO SWS, was revealed through
a careful modeling of the superimposed narrow CO gas-phase lines (Moneti, Cernicharo \& Pardo 2001).
Chiar et al. (2002) find that it is unlikely that the feature in the IRS~19 
spectrum is associated with the 
background illuminating source, given that it is an M supergiant with no obvious 
association with the local molecular cloud material.

In the central few arcseconds of the Galaxy the local FUV and X-ray radiation 
due to the hot stars and SgrA* in addition to the strong winds from these sources 
is likely responsible for the observed XCN absorption feature.

An alternative explanation for the 4.62$\mu$m absorption towards IRS~7
may be that it is a XCN feature that originates in a small line-of-sight clump
that happens to line up with IRS 7. Some of these clumps can be 
found especially close to the mini-spiral (see e.g. figure 14 in Eckart et al. 2006).

\subsection{Mass estimates of the stellar dust envelopes}\label{massestim}

With the exception of the hot IRS~16 sources and IRS2L all of the objects are dust embedded and
most of them are verified or likely bow shock sources.
Here we estimate the gas and dust masses of these 
sources and discuss the expected source intrinsic
M-band CO absorption.

\subsubsection{Mass estimate based on the mean mass density in the northern spiral arm}
Based on 63$\mu$m [OI] $^3$P$_1$$\rightarrow$$^3$P$_2$ emission line measurements,
Jackson et al. (1993) find a total of 300~\solm of neutral atomic hydrogen 
within the 1.5~pc radius circum nuclear ring. The dominating amount of that emission is located
along the northern spiral arm. With an overall extend of the northern arm of about 40''$\times$5''
this results in mean mass surface density of $\sim$1.5 \solm/arcsec$^2$.
From the 22'' resolution map of the [OI] line emission
shown in their Fig.5 we conclude that no more than about 1/5 of the material traced by 
the  63$\mu$m [OI] line is located along the inner 20'' section 
of the northern arm. 
From the analysis of the [OI] line emission the authors 
deduce a mean combined HI and H$_2$ particle density of (1-3)$\times$10$^5$~cm$^{-3}$.

Bradford et al. (2005) mapped the $^{12}$CO(7-6) line emission within 
the circum-nuclear ring.
A part of the molecular line emission  that can be attributed to the northern 
arm covers mostly the inner 20'' portion of it. 
For the northern arm Stacey et al. (2004) derive a gas mass of 5-50 \solm.
From the analysis of the CO(7-6) line emission the authors 
deduce a mean combined HI and H$_2$ particle density of 3$\times$10$^4$~cm$^{-3}$.
This value is consistent with the one used for bow shock modeling of these 
sources by Tanner et al. (2002, 2005).

From the mean particle density $n_{HI,H_2}$ (per cm$^{3}$) and the mean diameter of the sources of 0.1''
follows a mean gas mass of the gas and dust shell of about 
$M_{gas}[M_{\odot}]=1.7 \times 10^{-9} n_{HI,H_2} \eta \zeta$.
Here $\eta$ is an area filling factor (in projection on the sky) of the mini-spiral that we assume to be
of the order of 10\%.
The pile-up of mini-spiral material due to the interaction with the bow shock is described
by the factor $\zeta$. A bow shock source has typically moved half way through an 
approximately 4'' long stretch of the mini-spiral. The bow shock typically 
extends over about half a 0.1'' diameter sphere. In this scenario the bow shock may contain about
$\zeta$=4''/0.1''=10 times more gas and dust than the surrounding mini-spiral ISM.
The actual enhancement of the flux density in the bow shock is probably much larger
due to scattering of light from the central illuminating star. The amount of scattered light
depends on the grain size spectrum (see comments on IRS~21 in Tanner et al. 2002).
For densities in the range of 3$\times$10$^4$cm$^{-3}$$\le$$n_{HI,H_2}$$\le$3$\times$10$^5$cm$^{-3}$ we obtain
total gas masses of a few 10$^{-3}$\solm up to 10$^{-2}$\solm.
These masses could be larger by about an order of magnitude if the 
fractional abundances of CO versus H$_2$ for the diffuse ISM of 3$\times$10$^{-6}$ is used.
Since the mini-spiral gas densities derived from sub-mm CO(7-6) and FIR [OI] line data are 
typical for the dense interstellar medium, this would not be appropriate.

For the dust embedded hot stars like IRS~1W, IRS~10W and IRS~21 we find that
due to their expected fast stellar winds (mass losses of 10$^{-4}$ \solm
yr$^{-1}$ at a velocity of typically 1000~km/s;
Najarro et al. 1997 Tanner et al. 2002, 2005)
the kinetic luminosity of that wind
is comparable to that of the bow shock mass load due to the motion of the star
through the mini-spiral
(i.e. 10$^{-3}$-10$^{-2}$\solm at a velocity of 200-300 km/s over 10 to 20 years
travel time through the northern arm).
This leads to the fact that a bow shock is formed. If either of the two kinetic
luminosities is
much more dominant than the other then either a large cavity would be blown into the
mini-spiral or the motion 
of the star through the mini-spiral would not lead to a recognizable interaction.

\subsubsection{Typical absorption depths expected for the stellar dust envelopes}
In this section we calculate the typical source intrinsic 4.666$\mu$m CO gas line absorption
depths expected for the stellar dust envelopes 
assuming total gas masses of a few 10$^{-3}$\solm up to 10$^{-2}$\solm. 
We derive optical depths of the $^{13}$CO R(0) lines for the different observed sources from equation~\ref{equ1}.
For our calculations, we derived the H$_2$ column density by dividing the gas mass estimates by the sky projected area of the individual sources. Motivated by the brightness profiles for individual sources as given by Tanner et al. (2005) 
and taking into account the irregular horseshoe like shape of the bow shocks we doubled
the stand-off radius estimates in order to include all of the material. 
Then, given the densities suggested by the CO(7-6) and [OI] lines we have used
fractional abundances of $^{13}$CO versus H$_2$ of 5$\times$10$^{-6}$ in order to derive gaseous $^{13}$CO column densities.\\
On the other hand, we consider that the gas and dust in the shells and bow shocks is likely to be warm. Measurements and model calculations of the circum stellar shells in 
IRS~21, IRS~1W, IRS~5, IRS~10W by Tanner et al. (2002,2005) 
and IRS~3 by Pott et al. (2008) indicated dust temperatures 
between 400~K and 1000~K. 
Moreover, the 10$\mu$m dust temperature map by 
Cotera et al. (1999) yield averaged dust temperatures 
of 200~K towards the mini-spiral and most of the compact 
dust embedded sources.
\\
Besides, we assume that the random motions $\delta$$v$ within the stellar 
winds and bow shock regions are of the order 
of 15 to 25~km/s (Bradford et al. 2005, Eyres et al. 2004) 
possibly up to $\sim$100~km/s (see spectra in Fig.5 of Jackson et al. 1993). 
The FWHM of $\sim$100~km/s towards the individual sources obtained 
by Geballe, Baas \& Wade (1989) is most likely due to the velocity 
difference of the individual line of sight clouds. 

We find that if the gas masses of the stellar shells or bow shocks 
are of the order of a few times 10$^{-3}$-10$^{-2}$\solm and if we use doppler line widths of 25~km/s$<$$\delta$$v$$<$100~km/s
then we obtain $^{13}$CO optical depths of the local gas of $\tau_c\sim0.07-0.26$. At our spectral resolution of 375~km/s, we get optical depths of the order of $\tau_c\sim0.02$ which is in the range of the corrected optical depths for foreground extinction listed in Table~1. In that table, we give the observed optical depths $\tau_0$ of the $^{13}$CO R(0) derived from the spectra shown in Fig.~\ref{extcorr} of all observed sources at a resolution of 375~km/s. The actual optical depth at the line center at higher spectral resolutions may be higher. The $\tau_c$ values are the optical depths corrected for the mean foreground $^{13}$CO gas absorption 
as determined from the spectrum of the non-embedded hot IRS16C star ($\tau$$_{IRS16C}$$\sim$0.11$\pm$0.02, see section~\ref{The-gaseous-CO-absorption}). The absorption depths towards all other sources are higher
and can be determined with a correspondingly smaller uncertainty.
Except for IRS 3, 9, 5 13, 21, and 29 the optical depths are 2 to 3
sigma higher than that obtained towards IRS16C. \\
This indicates that even though the 4.666$\mu$m CO gas line 
absorption contains contributions from the intervening spiral arms,
local molecular gas and clouds close to the Galactic Center, 
a substantial amount of absorption observed at high spectral resolution 
can be expected to be due to the circumstellar shells of the 
dust embedded sources (see the case of IRS~3 below).

\subsubsection{Comparison to high spectral resolution observations}

This finding can be discussed in the context of the 
Fabry-Perot spectroscopic measurements by 
Geballe, Baas \& Wade (1989).
These spectra were obtained through 5" diameter apertures; 3.5" in the case of IRS~7.
In all cases these large apertures contain flux density contributions from several 
of the bright M-band sources that we study here.
These apertures are also ideally matched to the extended mini-spiral which is 
also bright in this wavelength domain.
Furthermore the pointing toward individual sources discussed by
Geballe, Baas \& Wade (1989) was carried out by offsetting from a visible star and 
'in most of the cases by peaking up on the infrared signal'.
No accuracy is given for this process.
In summary these individual measurements are sensitive to absorption 
covering about 20 square arcseconds each. Compared to the 0.6'' diameter slit
in seeing as good as 0.7'' this represents a dilution of the
absorption from the individual stars by up to a factor of 40.
The large area and the numerous possible absorbers 
may also be responsible for the width of the absorptions of 75-100~km/s.

For several sources observed by Geballe, Baas \& Wade (1989) radial
velocities are known: 
IRS~1W:  20$\pm$50 (Genzel et al. 2000);
IRS~5:  110$\pm$60 (Tanner et al. 2002, 2005);
IRS~7:  -103$\pm$15 (Genzel et al. 2000);
IRS~8:  +15$\pm$30 (Geballe et al. 2006).
For IRS~3 the findings by Geballe, Baas \& Wade (1989), 
Viehmann et al. (2005, 2006) and Pott et al. (2008)
indicate that it is not located within the central half parsec - implying low
radial velocities.
In fact, IRS~3 could be an example of a source for which the absorption due to
the circum-stellar shell may be of importance. The prominent J=1-0 R(2) absorption at 0~km/s 
is significantly weaker in the J=1-0 R(5) line indicating a colder or possibly lower column density
component (Geballe, Baas \& Wade 1989).
Since IRS~3 is a warm dusty source with a possibly low radial velocity. Therefore, the absorption line at 0 km/s could potentially be attributed to the gas and dust shell it is embedded in.

For IRS~1 the strong absorption at about 20~km/s reported by Geballe, Baas \& Wade (1989) 
falls close to the radial velocity of the source and could be an intrinsic absorption feature.
Similarly, the radial velocities of IRS~5, IRS~8, and possibly IRS~3 fall into 
or very close to the 0 - 75 km/s range where all of the sources show absorption.
Therefore the foreground and intrinsic source absorptions could blend with each other.
For IRS~7 Geballe, Baas \& Wade (1989) report intrinsic photospheric absorption with
a prominent feature at -130~km/s close to the radial velocity of
the star. However, if a substantial amount of absorbing material is associated with 
the IRS~7 bow-shock 
that high-lights the interaction of the stellar atmosphere with a wind from the
central region, then the source intrinsic absorption could also be closer to 
the 0 - 75 km/s interval and blend with the foreground absorption. This may explain the
slightly larger total line width seen towards that source.

With the exception of the early type stars IRS~13 and IRS~16NE the known radial velocities 
(e.g. Genzel et al. 2000, Blum et al. 2003 and Blum, Sellgren, Depoy, 1996)
of all other stars listed in Tab.~\ref{tau-table} should show intrinsic 
absorption features outside the 0 - 75 km/s interval.

\section{Summary}
We investigated the circumstellar material around 15 brightest MIR sources
- most of them are dust-enshrouded - in the central stellar cluster of the Milky Way.
We have presented $R=\Delta \lambda/\lambda$=800 ($\Delta v=375$~km/s) resolution M-band spectra taken in 
a $0.6\arcsec$ slit.
These spectra show a strong absorption centered at 4.666$\mu$m due to a mixture of
polar and apolar CO ice combined with a contribution of an XCN absorption feature at 4.62$\mu$m for some sources.
The CO ice absorption along the line of sight was derived after smoothing the IRS2L 
spectrum which is assumed to be free of local foreground extinction.
\\
We find that the extinctions due to material traced by the CO ice and the CO gas absorptions may be of comparable importance. Moreover, a major portion of the overall extinction can be explained 
as being due to the gas and dust of the cold neutral phase of the diffuse ISM.

Using the line of sight ice absorption spectrum and the one of IRS~16C we perform a first order correction 
of individual Galactic Center source spectra for the
foreground absorption due to CO-ice and gas, respectively. 
Based on published hydrogen number density estimates
from sub-mm CO(7-6) and FIR [OI] line data we 
derive gas masses of the circumstellar shells of the order of 10$^{-3}$\solm.
These estimates show that a substantial part of the 4.666$\mu$ gaseous CO absorption could be due to
intrinsic circumstellar material - in addition to absorption along the line of sight towards the center of the
Milky Way.
Future high spectral 
($R= \lambda/\Delta \lambda$$\sim$30000, i.e. $\sim$10~km/s)
and spatial resolution ($<$1'') are required to identify 
source intrinsic absorption
with a minimum of dilution by absorption features obtained in apertures of several arcseconds diameter.

\acknowledgments

This work was supported in part by the Deutsche Forschungsgemeinschaft 
(DFG) via grant SFB 494. 
We are grateful to all members of the ISAAC VLT team and to F. Najaro for useful discussions. We would like to thank the referee for his/her very helpful comments.



{\it Facilities:} \facility{VLT:Antu (ISAAC)}.

\bibliographystyle{apj}


\rf{Acharyya, K.; Fuchs, G. W.; Fraser, H. J.; van Dishoeck, E. F.; Linnartz, H., 2007, A\&A 466, 1005}

\rf{Allamandola, L. J., 1984, ASSL, 108, 5}
%
%
%
%
%

\rf{Bernstein, M.P.; Sandford, S.A.; Allamandola, L.J.; Chang, S.; Scharberg, M.A., 1995, ApJ 454, 327}

\rf{Bisschop, S. E.; Fraser, H. J.; \"Oberg, K. I.; van Dishoeck, E. F.; Schlemmer, S.	2006, A\&A 449, 1297}

\rf{Bohlin, R. C.; Savage, B. D.; Drake, J. F., 1978, ApJ 224,132}

\rf{Bradford, C. M.; Stacey, G. J.; Nikola, T.; Bolatto, A. D.; Jackson, J. M.; Savage, M. L.; Davidson, J. A., 2005, ApJ 623, 866}
 
 
%
%
%
%
%
%
%
%
%
\rf{Burgh, E.B., France, K., McCandliss, S.R.,2007 ApJ 658, 446}



%
\rf{Chiar, J. E.; Gerakines, P. A.; Whittet, D. C. B.; Pendleton, Y. J.; Tielens, A. G. G. M.; Adamson, A. J.; Boogert, A. C. A.,  1998, ApJ 498, 716}

%
%
\rf{Chiar, J. E., Adamson, A. J., Pendleton, Y. J., Whittet, D. C. B., 
    Caldwell, D. A., Gibb, E. L., 2002, ApJ 570, 198 }
 
\rf{Cl\'enet, Y., Rouan, D., Gendron, E., Montri, J.,
 Rigaut, F., L\'na, P., Lacombe, F. 2001, A\&A, 376, 124 }
 
%
%
%
%
\rf{Collings, M. P.; Dever, J. W.; Fraser, H. J.; McCoustra, M. R. S.; Williams, D. A.	2003, ApJ 583, 1058}

\rf{Cotera, A., Morris, M., Ghez, A. M., Becklin, E. E., Tanner, A. M., Werner, M. W., Stolovy, S. R. 1999, cpg conf, 240 }
%
%
%

\rf{Duley, W. W., 1974, Ap\&SS 26, 199}

\rf{Eckart, A.; Baganoff, F. K.; Sch\"odel, R.; Morris, M.; Genzel, R.; Bower, G. C.; 
Marrone, D.; Moran, J. M.; Viehmann, T.; Bautz, M. W.; and 10 coauthors, 2006, A\&A 450, 535}
 
 
%
%
 
 
%
%
%
\rf{Eyres, S. P. S.; Geballe, T. R.; Tyne, V. H.; Evans, A.; Smalley, B.; Worters, H. L., 2004, MNRAS, 350, L9}
%
%
%
%
\rf{Ferriere, K.M.; Zweibel, E.G.; Shull, J.M., 1988, ApJ 332, 984}

\rf{Field, G.B.; Goldsmith, D.W.; Habing, H.J., 1969, ApJ 155, L149}

\rf{Fong, D.; Meixner, M.; Castro-Carrizo, A.; Bujarrabal, V.; Latter, W. B.; Tielens, A. G. G. M.; Kelly, D. M.; Sutton, E. C., 2001, A\&A 367, 652}

\rf{Geballe, T.R., Wollman, E.R., \& Rank, D.M., 1972, ApJ 177, L27}
 
\rf{Geballe, T.R., 1986, A\&A 162, 248}

\rf{Geballe,T.R.; Baas, F. and Wade, R., 1989, A\&A 208, 255} 

\rf{Geballe, T. R.; Najarro, F.; Rigaut, F.; Roy, J.-R., 2006, ApJ 652, 370}

\rf{Genzel, R., Thatte, N., Krabbe, A., Kroker, H., Tacconi-Garman, L. E. 1996, ApJ 472, 153 }
 
%
\rf{Genzel, R., Pichon, C., Eckart, A., Gerhard, O. \& Ott, T. 2000,
 Mon.Not.R.Soc.317, 348-374 }
\rf{Guesten, R., Genzel, R., Wright, M. C. H., Jaffe, D. T., Stutzki, J., Harris, A. I. 1987, ApJ 318, 124 }
%
%
\rf{Hagen, W.; Allamandola, L. J.; Greenberg, J. M; 1979, Ap\&SS, 65, 215}

\rf{Hagen, W.; Allamandola, L. J.; Greenberg, J. M; 1980 A\&A, 86, L3}

\rf{Hippelein, H.; Haas, M.; Tuffs, R. J.; Lemke, D.; Stickel, M.; Klaas, U.; V\"o, H. J., 2003, A\&A 407, 137}

%
%
\rf{Hotzel, S., Harju, J., Juvela, M., Mattila, K., Haikala, L. K., 2002, A\&A 391,
275}

%
%
%
\rf{Jackson, J. M.; Geis, N.; Genzel, R.; Harris, A. I.; Madden, S.; Poglitsch, A.; Stacey, G. J.; Townes, C. H., 1993, ApJ 402, 173 }
%
%
\rf{Jones, A. P.,  2002, EAS Publications Series, Volume 4, 
Proceedings of Infrared and Submillimeter Space Astronomy", 
held 11-13 June, 2001. Edited by M. Giard, J.P. Bernanrd, 
A. Klotz, and I. Ristorcelli. EDP Sciences, 2002, pp.37-37}

%
%
%
%

\rf{Kainulainen, J., Lehtinen, K., Harju, J., 2006, A\&A 447, 597}
\rf{Klaassen, P. D., Plume, R., Gibson, S. J., Taylor, A. R., Brunt, C. M., 2005,
ApJ 631, 1001}

\rf{Krabbe, A. et al. 1995, ApJL 447, L95 }
 

\rf{Lacy, J.H.; Baas, F.; Allamandola, L.J.; van de Bult, C.E.P.; Persson, S.E.; McGregor, P.J.; Lonsdale, C.J.; Geballe, T.R., 1984, ApJ 276, 533}

\rf{Lacy, J. H. \& Achtermann, J. M. 1991, ApJ 380, L71-L74}

\rf{Langer, W.D., Penzias, A.A., 1990, ApJL 357, 477}

\rf{Lebofsky, M. J.; Rieke, G. H.; Tokunaga, A. T., 1982, ApJ 263, 736L}
 
%
\rf{Lee, T. J., 1972, Nature 237, 99L}
\rf{Lee, T. J., 1975, Ap\&SS,  34, 123L}
\rf{Liszt, H.S., 2007, A\&A Letters 476, 291}

%
%
%
%
\rf{McFadzean, A.D., Whittet, D.C.B., Bode, M.F., Adamson, A.J., Longmore, A.J. 1989, MNRAS 241, 873 }

\rf{Mellema, G.; Raga, A. C.; Canto, J.; Lundqvist, P.; Balick, B.; Steffen, W.; Noriega-Crespo, A., 1998, A\&A 331, 335}

\rf{Mennella, V., Palumbo, M. E., Baratta, G. A., 2004, ApJ 615, 1073}

%

\rf{Millar, T. J.; Freeman, A., 1984a, MNRAS 207, 405}

\rf{Millar, T. J.; Freeman, A.  1984b, MNRAS 207, 425}

\rf{Moneti, A., Cernicharo, J., Pardo, J.R. 2001, ApJ 549, L203 }

\rf{Morris, M., \& Maillard, J.-P. 2000, in Imaging the Universe in Three Dimensions: Astrophysics with Advanced Multi-Wavelenght Imaging Devices., ed. W. van Breugel, \& J. Bland-Hawthorn, ASP Conf. Ser., 195, 196 }

\rf{Moultaka, J.; Eckart, A.; Sch\"odel, R.; Viehmann, T.; Najarro, F., 2005, A\&A 443, 163}

\rf{Moultaka, J.; Eckart, A.; Viehmann, T.; Mouawad, N.; Straubmeier, C.; Ott, T.; Sch\"odel, R., 2004, A\&A 425, 529}
 
\rf{Muzic, K; Eckart, A.; Sch\"odel, R.; Meyer, L.; Zensus, A.;  2007; A\&A, 469,993}

\rf{Muzic K.; Sch\"odel, R.; Eckart, A.; Meyer, L.; Zensus, A., 2008, A\&A 482, 173}

%
\rf{Najarro, F., Krabbe, A., Genzel, R., Lutz, D., Kudritzki, R. P., Hillier, D. J. 1997, A\&A 325, 700 }
%

\rf{Neininger, N.; Guelin, M.; Garcia-Burillo, S.; Zylka, R.; Wielebinski, R.,
1996, A\&A 310, 725}

\rf{ O'Neill, P. T.; Viti, S.; Williams, D. A.  2002, A\&A 388, 346O}

\rf{Ott, T.; Eckart, A.; Genzel, R., 1999, ApJ 523, 248}

%



\rf{Paumard, T., Maillard, J.-P., Morris, M., \& Rigaut, F. 2001, A\&A, 366, 466 }

\rf{Paumard, T.; Genzel, R.; Martins, F.; Nayakshin, S.; Beloborodov, A. M.; Levin, Y.; Trippe, S.; Eisenhauer, F. et al. 2006, ApJ, 643, 1011}

%
\rf{Pendleton, Y. J.; Tielens, A. G. G. M.; Tokunaga, A. T.; Bernstein, M. P.,  1999, ApJ 513, 294}
%

\rf{Pott, J.-U.; Eckart, A.; Glindemann, A.; SchÃ¶del, R.; Viehmann, T.; Robberto, M,  2008, A\&A 480, 115}

\rf{Rodriguez-Fernandez, N. J.; Martin-Pintado, J.; Fuente, A.; de Vicente, P.; Wilson, T. L.; Huettemeister, S., 2001, A\&A 365, 174}
%
%
%
%
%
%
%
%
\rf{Sandford, S. A.; Allamandola, L. J.; Tielens, A. G. G. M.; Valero, G. J., 1988, ApJ 329, 498}

%
%
\rf{Scoville, N.Z., Stolovy, S.R., Rieke, M., Christopher, M.H., Yusef-Zadeh F. 2003 ApJ, 594, 294 }

\rf{Schutte, W.A.; Greenberg, J.M., 1997, A\&A 317, L43}

\rf{Sch\"odel, R.; Eckart, A.; Alexander, T.; Merritt, D.; Genzel, R.; Sternberg, A.; Meyer, L.; 
    Kul, F.; Moultaka, J.; Ott, T.; Straubmeier, C., 2007, A\&A 469, 125}
 
 
%
%
%
%
%
\rf{Sonnentrucker, P., Welty, D. E., Thorburn, J. A., York, D. G., 2007, ApJS 168, 58}

\rf{Suzuki, T.; Kaneda, H.; Nakagawa, T.; Makiuti, S.; Okada, Y.; Shibai, H.; Kawada, M., 2009,
The Evolving ISM in the Milky Way and Nearby Galaxies, The Fourth Spitzer Science Center 
Conference,  p.63, Proceedings of the conference held December 2-5, 2007 at the Hilton Hotel, 
Pasadena, CA, Eds.: K. Sheth, A. Noriega-Crespo, J. Ingalls, and R. Paladini, 
Published online at http://ssc.spitzer.caltech.edu/mtgs/ismevol/}

%
%
%
%
\rf{Tanner, A., Ghez, A. M., Morris, M., Becklin, E. E., Cotera, A., 
    Ressler, M., Werner, M., Wizinowich, P. 2002, ApJ 575, 860}
 

\rf{Tanner, A., Ghez, A. M., Morris, M., Christou, J. C., 2005, ApJ 624, 742}

\rf{Tegler, Stephen C.; Weintraub, David A.; Allamandola, Louis J.; Sandford, Scott A.; Rettig, Terrence W.; Campins, Humberto, 1993, ApJ 411, 260} 

\rf{Tegler, Stephen C.; Weintraub, David A.; Rettig, Terrence W.; Pendleton, Yvonne J.; Whittet, Douglas C. B.; Kulesa, Craig A., 1995, ApJ 439, 279}

\rf{Teixeira, T. C.; Emerson, J. P., 1999, A\&A 351, 292}

%
%
%
\rf{Tielens, A. G. G. M.; Tokunaga, A. T.; Geballe, T. R.; Baas, F.; 1991, ApJ, 381, 181}
%
%
%

\rf{Van Dishoeck, E.F and Black, J. H., 1986, ApJS 62, 109} 

\rf{van Loon, J. Th.; Oliveira, J. M., 2003, A\&A 405 , L33}

\rf{Viehmann, T., Eckart, A., Sch\"odel, R., Moultaka, J., Straubmeier, C., Pott, J.U. 2005, A\&A, 433, 117}

\rf{Viehmann, T.; Eckart, A.; Sch\"odel, R.; Pott, J.-U.; Moultaka, J.,  2006, ApJ 642, 861}

\rf{Viti, S.; Williams, D. A.; O'Neill, P. T.  2000, A\&A 354, 1062}

%

\rf{Walmsley, C. M., 1997, Int. Astron. Union Symp., No. 170, p. 79}

\rf{Whittet, D.C.B., Boogert, A.C.A, Gerakines, P.A. et al., 1997, ApJ 490, 729}
	
\rf{Wolfire, M.G.; Hollenbach, D.; McKee, C.F.; Tielens, A.G.G.M.; Bakes, E.L.O., 1995, ApJ 443, 152}
%
%
%

\rf{Yusef-Zadeh, F.; Morris, Mark, 1991, ApJ 371, L59}






\clearpage



\clearpage

\clearpage

\begin{table*}[htbp]
\small
\begin{center}
\begin{tabular}{lr}
\hline
Source       &  Value of the diluting     \\
       &  continuum (Jy)     \\
\hline
IRS16C     & 4 \\
IRS16NE    & 4 \\
IRS16SW    & 4 \\
IRS2S      & 3 \\
IRS2L      & 3 \\
IRS12N     & 4 \\
IRS9       & 3 \\
IRS3       & 5.5 \\
IRS1W      & 5 \\
IRS5       & 3 \\
IRS10W     & 4 \\
IRS13      & 4 \\
IRS29      & 4 \\
IRS7       & 3 \\
IRS21      & 2 \\
\hline
\end{tabular}
\end{center}
\caption{Values of the diluting continuum added to the solid CO ice absorption template shown in Fig.~\ref{icegas}(a) to correct the spectra of the GC sources for the corresponding foreground absorption.
The mean of 3.7$\pm$0.9 is rounded up to a value of 4.0 in the text.
}
\label{dilution}
\end{table*}

\begin{table}[htb]
\begin{center}
\begin{tabular}{ccc}\hline \hline
                                & Mean          & Mean  \\ 
                               & Diffuse       & Dense  \\ 
                               & Clouds        & Clouds  \\
\hline 
Typical              &      &  \\ 
$^{12}$CO            & -5.5          & -4.1  \\
abundance            &      &  \\
                                   &   &   \\
N(CO$_{ice}$;$\tau$=0.1)   & 16.61 & 16.61 \\
N(H$_2$)                   & 22.14 & 20.71  \\
A$_V$                       & 13.8 - 27.6   & 0.8  \\
                           &   &   \\
N(CO$_{ice}$;$\tau$=0.2)   & 16.91 & 16.91 \\
N(H$_2$)                   & 22.43 & 21.00  \\
A$_V$                      & 26.7 - 53.3   & 1.5  \\
\hline \hline
\end{tabular}
\end{center}
\caption{ A summary of the logarithmic H$_2$ abundances and optical extinctions A$_V$ from column densities of CO ice towards the Galactic Center. In the first line, we list the typical mean values of the $^{12}$CO abundance for the diffuse interstellar medium ([$^{12}$CO]/[H$_2$]=3$\times$10$^{-6}$)
and dense Galactic molecular clouds 
([$^{12}$CO]/[H$_2$]=8$\times$10$^{-5}$).
In the following lines are listed the CO ice column densities corresponding to the CO ice optical depths that we found of $\tau_{4.6\mu m}$=0.1-0.2. We also list the derived H$_2$ column density as well as the deduced
extinctions A$_V$.
The abundances are based on values derived or given in 
Millar \& Freeman (1984 a, b),
van Dishoeck and Black (1986),
Walmsley (1997),
Hotzel et al. (2002),
Burgh, France, McCandliss (2007),
Sonnentrucker et al. (2007),
Liszt (2007). The extinction calibration uses A$_V$/N(H$_2$) and A$_V$/N(CO) values towards a variety of sources from Bohlin et al. 1978,
Teixeira \& Emerson 1999, Mennella et al. 2004, Klaassen et al. 2005,
Kainulainen et al. 2006, Moneti, Cernicharo \& Pardo 2001. Two values are shown for the calculated $A_V$ corresponding to A$_V$/N(H$_2$) of about $1.- 2.~10^{-21}$.}
\label{abundance}
\end{table}

\clearpage

\begin{table*}[htbp]
\small
\begin{center}
\begin{tabular}{llrrrrrrrrrrrrrrr}
\hline
       &       & radial   &        &         &        &        &  \\
Source & type  & velocity &M-band & size    & T      &$\tau_0$ &$\tau_c$  \\
       &       & km/s     &mag.   & arcsec  & K      &         & \\
\hline
IRS16C    & hot  & 180 $\pm$25 & 7.78& 0.10& 400&   0.11 & 0.0\\
IRS16NE   & hot  &  17 $\pm$25 & 6.87& 0.10& 400&   0.15 & 0.04\\
IRS16SW   & hot  & 460 $\pm$30 & 7.53& 0.10& 400&   0.17 & 0.06\\
IRS2S     & cool & 107 $\pm$20 & 7.53& 0.10& 400&   0.15 & 0.04\\
IRS2L     & hot  & -           & 5.55& 0.10& 400&   0.15 & 0.04\\
IRS12N    & cool & -96 $\pm$20 & 6.43& 0.10& 400&   0.16 & 0.05\\
IRS9      & hot  & -300$\pm$25 & 6.90& 0.10& 400&   0.12 & 0.01\\
IRS3      & cool & -           & 3.35& 0.50& 450&   0.12 & 0.01\\
IRS1W     & hot  &  20 $\pm$50 & 4.20& 0.12& 400&   0.15 & 0.04\\
IRS5      & cool & 110 $\pm$60 & 4.90& 0.12& 400&   0.13 & 0.02\\
IRS10W    & cool &  80 $\pm$60 & 5.02& 0.10& 400&   0.15 & 0.04\\
IRS13     & hot  &  45 $\pm$60 & 5.73& 0.10& 400&   0.12 & 0.01\\
IRS29     & hot  & -93 $\pm$20 & 6.31& 0.10& 400&   0.14 & 0.03\\
IRS7      & cool & -103$\pm$15 & 4.30& 0.10& 400&   0.19 & 0.08\\
IRS21     & hot  & -90 $\pm$20 & 5.47& 0.18& 400&   0.14 & 0.03\\
\hline
\end{tabular}
\end{center}
\caption{Parameters and results for the molecular gas mass estimate of the
circumstellar shells and bow-shocks
of the 15 bright MIR sources that we observed in the Galactic center.
The identification of the stars as 'hot' (i.e. WC,WR or O-star) or 'cool' (giant,
super giant) is based on
e.g. 
Lebofsky et al. (1982),
Krabbe et al. (1995),
Genzel et al. (1996), (2000),
Ott et al. (1999),
Tanner et al. (2002, 2005), 
Paumard et al. (2006) and Cl\'enet et al. (2001).
The M-band magnitudes are from Viehmann et al. (2005, 2006).
The gas and dust temperatures estimates $T$ are based on  Cotera et al. (1999),
Tanner et al. (2002,2005),  Pott et al. (2008).
$\tau_0$ are the observed optical depths 
of the $^{13}$CO R(0) and $\tau_c$ are the corrected optical depths for 
the mean foreground $^{13}$CO gas absorption (see text). 
For references of radial velocity and classification: For IRS~1W, 5, 10W, 21, see Tanner et al. (2005), 
Genzel et al. (2000) for all others see Blum et al. (2003) and Blum et al. (1996).
}
\label{tau-table}
\end{table*}


\end{document}